\documentclass[twocolumn,showpacs,amsmath,amssymb,floatfix,pre]{revtex4}
\usepackage{graphicx}
\begin{document}
\title{Entropy of polydisperse chains: solution on the Bethe lattice} 
\author{Minos A. Neto}
\email{minos@if.uff.br}
\author{J\"urgen F. Stilck}
\email{jstilck@if.uff.br}
\affiliation{Instituto de F\'{\i}sica\\
Universidade Federal Fluminense\\
Av. Litor\^anea s/n\\
24210-346 - Niter\'oi, RJ\\
Brazil}
\date{\today}

\begin{abstract}
We consider the entropy of polydisperse chains placed on a
lattice. In particular, we study a model for equilibrium
polymerization, where the
polydispersivity is 
determined by two activities, for internal and endpoint monomers of a
chain. We solve the problem exactly on a Bethe lattice with arbitrary
coordination number, obtaining an expression for the entropy as a
function of the density of monomers and mean molecular weight of the
chains. We compare this entropy with the one for the monodisperse
case, and find that the excess of entropy due to polydispersivity is
identical to the one obtained for the one-dimensional case. Finally,
we obtain an exponential distribution of molecular weights. 
\end{abstract}

\pacs{65.50.+m,05.20.-y}

\maketitle

\section{Introduction}
\label{intro}
The problem of the thermodynamic behavior of diatomic molecules which
are adsorbed on two-dimensional surfaces has attracted the interest of
researchers already in the thirties \cite{fr37}. In the simplest model of
this kind, only excluded volume interactions are taken into account,
and the relevant quantity is the entropy of placing dimers (which
occupy two first neighboring sites) on the lattice, which is a
fundamental equation of the system. For the particular
case of full occupancy and two-dimensional lattices, this problem was
solved exactly by Kasteleyn and Temperley and Fisher \cite{k61}. The
model may be generalized in many ways. For example, an energy may be
associated to the configurations of dimers on the lattice, and such
models may display rather peculiar phase transitions
\cite{nyb89}. Also, we may generalize the athermal model allowing for
chains of more than two monomers ($M$-mers) and also considering only
a fraction $\rho$ of the sites of the lattice occupied by monomers. 

The entropy of $M$-mers (chains of molecular weight $M$), as a
function of the fraction of lattice sites 
occupied by monomers $\rho$, is a fundamental equation of this system,
and thus contains all thermodynamic information. The entropy may be
defined as
\begin{equation}
s_M(\rho)=\frac{1}{V}\lim_{V \to \infty} \ln \Gamma(N_p,M;V),
\end{equation}
where $\Gamma(N_p,M;V)$ is the number of ways to place $N_p$ chains with $m$
monomers in each on the lattice with $V$ sites, 
and the thermodynamic limit is taken with fixed density of occupied sites
$\rho=N_p M/V$. Besides the exact result mentioned above for $s_2(1)$
in two dimensions, in the literature we may find series expansions
estimates \cite{ncf89}, closed form approximations (Bethe and Husimi
lattices) \cite{so90} and transfer matrix calculations with
finite size extrapolations \cite{ds03} on the square lattice.

It is frequent in polymeric systems that the chains do not all have
the same number of monomers, that  is, a {\em polydisperse} set of
chains is found. One particular model for polydisperse chains is the
equilibrium polymerization model proposed by Wheeler, Kennedy and
Pfeuty \cite{wkp80,wp81} and applied to the study of equilibrium
polymerization of 
sulfur. The system is is defined on a lattice (celular model), so that
each of the $V$ cells is occupied by a monomer (a $\mathrm{S_8}$ ring
in the case of sulfur). Each monomer may be {\em active} or
{\em inactive} (open and closed rings for sulfur), and active
monomers in adjacent sites may connect, forming linear polymers which
are self-and mutually avoiding walks. Active monomers which are not
connected to any other ones are considered one-site polymers. The
statistical weight of chain composed by $m$ monomers (and therefore by
$m-1$ bonds) is given by
\begin{eqnarray}
K_1 & \mbox{if $M=1$,}\\
2K_1(K_p^\prime)^{M-1} & \mbox{if $M>1$.}
\end{eqnarray}
The additional factor 2 in the weight of multi-site chains is
necessary for the exact correspondence of the equilibrium
polymerization model to the $n$-vector of magnetism in the formal limit
$n \to 0$, but it may also be justified by a combinatorial argument
\cite{wp81}. This model, as well as related ones, was studied 
in some detail 
in the literature \cite{dfd99}, and experimental realizations of this
situation are found in equilibrium polymerization of sulfur and in
the so called living polymers \cite{g98,pw83}. It is expected that for the
polydisperse case the entropy is higher as compared to the
monodisperse case, due to the additional degrees
of freedom. Another relevant point is the distribution of molecular
weights $M$. Recently the entropy of polydisperse chains and the
distribution of molecular weights of the chains were calculated in the
one-dimensional case \cite{snd06}. In this paper we address these
questions for the solution of the model on the Bethe lattice with
coordination number $q$. As in the one-dimensional case, on the Bethe
lattice it is also possible to reach closed form expressions for the
entropy and for the distribution of molecular weights.

I the section \ref{mods} the problem is defined in more detail and the
relation of the equilibrium polymerization model to a simpler version,
without the presence of one-site polymers, is shown. The
entropy and the distribution of molecular weights on the Bethe lattice
is obtained. Final discussions and comments may be found in section
\ref{fdc} 

\section{Definition of the model and solution on the Bethe lattice}
\label{mods}
\begin{figure}[h]
\includegraphics[scale=0.6]{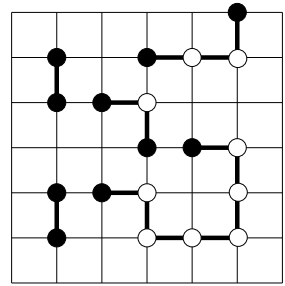}
\caption{A configuration of chains placed on the lattice. Internal
  monomers are represented as white circles and endpoint monomers are
  black circles. The statistical weight of this configuration with 5
  chains is $z_i^9\,z_e^{10}$.} 
\label{f1}
\end{figure}

The partition function for model for equilibrium polymerization
described above may be written as:
\begin{equation}
Y(K_1,K_p,x_1;V)=\sum (2K_1)^{N_p}(K_p^\prime)^{N_b}x_1^{N_1},
\end{equation}
where the sum is over all possible configurations of the polymers on
the lattice, $N_p$ is the number of polymers (including one-site
chains) of the configuration, $N_b$ corresponds to the number of bonds
and $x_1=1/2$ is a factor which assures the proper counting of
configurations, as discussed above. We may now perform a partial sum
over configurations in this partition function. Let us consider a
particular configuration of the polymers which occupy more than one
site ($M>1$). The number of monomers in internal site of the chains
will be called $N_i$ and the number of the monomers on endpoints of
the chains is $N_e$. The number of bonds in this configuration is
$N_b=N_i+N_e/2$ the total number of chains is $N_p=N_1+N_e/2$ and the
number of chains with more than one monomer is 
equal to $N_e/2$. We may now write the partition function as:
\begin{eqnarray}
Y(K_1,K_p,x_1;V)=\nonumber \\
{\sum}^\prime (2K_1)^{N_e/2}(K_p^\prime)^{N_e/2+N_i}
{\sum}^{\prime\prime}(2K_1x_1)^{N_1},
\end{eqnarray}
where the first sum is restricted to configurations of multi-site
chains and the second sum is over the configurations of the one-site
polymers for a fixed configuration of the other chains. It is now
trivial to perform the second sum, which is equal to
$(1+2K_1x_1)^{V-N_i-N_e}$, where we notice that $V-N_i-N_e$ is the
number of sites not occupied by monomers in multi-site chains and each
of these sites may be either empty (weight 1) or occupied by a
one-site polymer (weight $2K_1x_1$). Thus, the partition function of
the equilibrium polymerization model may be written as:
\begin{eqnarray}
Y(K_1,K_p,x_1;V)&=&(1+2K_1x_1)^V {\sum}^\prime z_e^{N_e}z_i^{N_i}
\nonumber\\
&=&(1+2K_1x_1)^V \Xi(z_e,z_i;V),
\label{yxi}
\end{eqnarray}
where the sum over configurations is now restricted to chains with
more than one monomers, the 
variable $x_1$ is equal to $1/2$ and the activities of endpoint and
internal monomers are
\begin{equation}
z_e=\frac{\sqrt{2K_1 K_p^\prime}}{1+2K_1x_1}
\label{ze}
\end{equation}
and
\begin{equation}
z_i=\frac{K_p^\prime}{1+2K_1x_1},
\label{zi}
\end{equation}
respectively. We may now restrict our discussion to the model without
one-site monomers, described by the grand-canonical partition function
$\Xi(z_e,z_i;V)$. 
In Fig. \ref{f1} a possible configuration of the chains is
shown. 
If the lattice has $V$ sites, the density of endpoint monomers is
$\rho_e=N_e/V$, the density of internal monomers is $\rho_i=N_i/V$,
and the total density of monomers is $\rho=\rho_e+\rho_i$. The
densities may be obtained from the partition function:
\begin{equation}
\rho_i=\frac{z_i}{\Xi}\frac{\partial{\Xi}}{\partial z_i},
\label{ri}
\end{equation}
and
\begin{equation}
\rho_e=\frac{z_e}{\Xi}\frac{\partial{\Xi}}{\partial z_e}.
\label{re}
\end{equation}

\begin{figure}[h]
\includegraphics[scale=0.6]{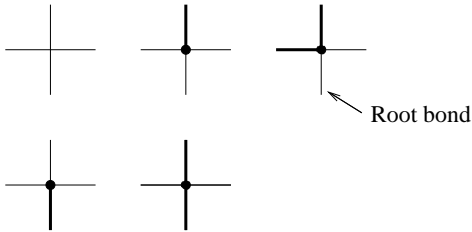}
\caption{Contributions to the recursion relations for the partial
  partition functions. The first line are the contributions to
  $g^{\prime}_0$ and the second line shows the contributions to
  $g^{\prime}_1$.} 
\label{f2}
\end{figure}

Let us now consider the model defined on a Cayley tree with
coordination number $q$. Due to the hierarchical structure of this
lattice, we may define partial partition functions on rooted subtrees,
and it is not difficult to write down recursion relations for subtrees
with an additional generation \cite{b82}. For the present model, we
may define two partial partition functions $g_0$ and $g_1$, without
and with a polymer bond on the root bond of the subtree,
respectively. Considering the operation of attaching $\sigma=q-1$
$n$-generations subtrees to a new site and bond, we obtain recursion
relations for the partial partition functions of a subtree with $n+1$
generations. The contributions to this recursion relations are shown
in Fig. \ref{f2}, and the resulting expressions are
\begin{eqnarray}
g_0^{\prime}&=&g_0^\sigma+\sigma z_e
g_0^{\sigma-1}g_1+\frac{\sigma(\sigma-1)}{2}z_ig_0^{\sigma-2}g_1^2. \\
g_1^{\prime}&=&z_e g_0^\sigma +\sigma z_i g_0^{\sigma-1} g_1.
\end{eqnarray}
If we now define the ratio of partial partition functions $R=g_1/g_0$,
we will find the following recursion relation:
\begin{equation}
R^{\prime}=\frac{z_e+\sigma z_i R}{1+\sigma z_e R
  +\frac{\sigma(\sigma-1)}{2} z_i R^2}.
\label{rrr}
\end{equation}
In the thermodynamic limit, the fixed point of this recursion relation
is reached. 

The partition function of the model on the Cayley tree may be obtained
if we consider the operation of attaching $q$ subtrees to the central
site of the tree. The result is:
\begin{equation}
\Xi=g_0^q + q z_e g_0^{q-1} g_1+\frac{q(q-1)}{2} z_i g_0^{q-2} g_1^2.
\end{equation}
The Bethe lattice corresponds to the central region of the Cayley
tree, and therefore we concentrate our attention on the densities of
endpoint monomers and of internal monomers on this site. The results
are: 
\begin{equation}
\rho_e=\frac{q z_e R}{1+q z_e R +\frac{q(q-1)}{2} z_i R^2},
\label{rhoe}
\end{equation}
and
\begin{equation}
\rho_i=\frac{\frac{q(q-1)}{2} z_i R^2}{1+q z_e R +\frac{q(q-1)}{2} z_i
  R^2}.
\label{rhoi}
\end{equation}
For given values of the activities $z_e$ and $z_i$, the fixed point
value of the ratio $R$ may be found using Eq. (\ref{rrr}) and thus
the corresponding densities of endpoint and internal monomers at the
central site of the tree are defined by Eqs. (\ref{rhoe}) and
(\ref{rhoi}), respectively. For $z_e>0$, the densities are smooth
functions of the activities, but in the polymer limit (infinite
chain) $z_e \to 0$ a continuous phase transition is found between a
non-polymerized phase $\rho_i=0$ and a polymerized phase $\rho_i>0$,
at $z_i=1/(q-1)$ \cite{g84,sw87}.

If we call $s(\rho_e,\rho_i)$ the entropy per site of the model, we
notice that the ratios of the chemical potentials and $k_BT$ are given
by 
\begin{equation}
\frac{\mu_e}{k_BT}=\ln z_e=-\left( \frac{\partial s}{\partial \rho_e}
\right)_{\rho_i} ,
\label{se}
\end{equation}
and
\begin{equation}
\frac{\mu_i}{k_BT}=\ln z_i=-\left(\frac{\partial s}{\partial \rho_i}
\right)_{\rho_e}. 
\label{si}
\end{equation} 
We then may obtain the entropy integrating the chemical potentials. To
do this, we start inverting equations for the densities (\ref{rhoe}) and
(\ref{rhoi}), which leads to:
\begin{equation}
z_e=\frac{\rho_e}{q(1-\rho_i-\rho_e)R},
\label{zer}
\end{equation}
and
\begin{equation}
z_i=\frac{2\rho_i}{q(q-1)(1-\rho_i-\rho_e)R^2}.
\label{zir}
\end{equation}
If these activities are now substituted into the fixed point equation
$R^{\prime}=R$, we obtain, from Eq. (\ref{rrr}) the fixed point
value of the ratio:
\begin{equation}
R^2=\frac{2\rho_i+\rho_e}{q-2\rho_i-\rho_e}.
\label{fpr}
\end{equation}
Substituting this result in Eqs. (\ref{zer}) and (\ref{zir}), we
obtain 
the activities as functions of the densities in the thermodynamic
limit:
\begin{equation}
z_e=\frac{\rho_e \sqrt{q-2\rho_i-\rho_e}}
  {q(1-\rho_i-\rho_e)\sqrt{2\rho_i+\rho_e}},
\end{equation}
and
\begin{equation}
z_i=\frac{2\rho_i(q-2\rho_i-\rho_e)}{q(q-1)(1-\rho_i-\rho_e)
  (2\rho_i+\rho_e)}.
\end{equation}
The entropy may now be calculated integrating the Eqs. (\ref{se})
and (\ref{si}) in the $(\rho_e,\rho_i)$ plane. We choose the trajectory
$(0,0) \to (0,\rho_i) \to (\rho_e,\rho_i)$, so that:
\begin{eqnarray}
s(\rho_e,\rho_i)&=&-\int_0^{\rho_i} \ln \left[
  \frac{q-2\rho}{q(q-1)(1-\rho)} \right] d\rho -\nonumber \\
&&\int_0^{\rho_e} \ln \left[\frac{\rho\sqrt{q-2\rho_i-\rho}}
  {q(1-\rho_i-\rho)\sqrt{2\rho_i+\rho}} \right] d\rho.
\end{eqnarray}
The result of this integration is:
\begin{eqnarray}
s(\rho_e,\rho_i)&=&-\frac{1}{2}(q-2\rho_i)\ln q +\rho_i \ln (q-1)-\rho_e
\ln \frac{\rho_e}{q}+ \nonumber \\
&&\frac{1}{2}(q-2\rho_i-\rho_e) \ln (q-2\rho_i-\rho_e) - \nonumber \\
&&(1-\rho_i-\rho_e) \ln (1-\rho_i-\rho_e) + \nonumber \\
&&\frac{1}{2} (2\rho_i+\rho_e)\ln (2\rho_i+\rho_e)- \rho_i \ln
(2\rho_i).
\label{srhoie}
\end{eqnarray}

The entropy may be compared to results in the literature
in some particular cases. For $\rho_e=0$ the case of infinite polymers
is obtained, and the expression reduces to Eq. (26) in reference
\cite{so90}. The entropy of dimers on the Bethe lattice (Eq. (27) in
reference \cite{so90}, where in the correct result $\rho z$ should be
replaced by $\rho/z$) is obtained for $\rho_i=0$. Finally, the
one-dimensional result (Eq. (9) in reference \cite{snd06}) is
recovered for $q=2$. To compare the result with the monodisperse case,
it is convenient to rewrite the entropy Eq. (\ref{srhoie}) as a function of
the mean molecular weight $\bar{M}=2(1+\rho_i/\rho_e)$ and the total
density of monomers $\rho=\rho_i+\rho_e$. The resulting expression is:
\begin{eqnarray}
s_{\bar{M}}(\rho)&=&\frac{\rho}{\bar{M}} \ln q+\rho
\frac{\bar{M}-2}{\bar{M}}\ln (q-1) - 
\frac{2\rho}{\bar{M}}\ln \frac{2\rho}{\bar{M}}+\nonumber \\
&&\left(\frac{q}{2}-\rho\frac{\bar{M}-1}{\bar{M}} \right)
\ln \left( 1-2\rho\frac{\bar{M}-1}{q\bar{M}} \right)- \nonumber \\
&&(1-\rho)\ln (1-\rho)+\rho\frac{\bar{M}-1}{\bar{M}}\ln \left(
2 \rho\frac{\bar{M}-1}{\bar{M}} \right) - \nonumber \\
&&\rho\frac{\bar{M}-2}{\bar{M}} \ln \left(
\rho\frac{\bar{M}-2}{\bar{M}} \right).
\end{eqnarray}
We may now find the contribution to the entropy of polydispersivity,
calculating $\Delta s_M(\rho)=s_{\bar{M}}(\rho)-s_M(\rho)$ for
$\bar{M}=M$, where $s_M(\rho)$ is the entropy of the monodisperse case
(Eq. (22) of reference \cite{so90}). We find a result which is
independent of $q$:
\begin{equation}
\Delta s_M(\rho)=\frac{\rho}{M}[(M-1)\ln(M-1)-(M-2)\ln(M-2)],
\end{equation}
and therefore this result is identical to the one obtained in the
one-dimensional case studied in reference \cite{snd06}.

Finally, we may find the distribution of the molecular weights in the
polydisperse case. To obtain this result, we must keep track of the
number of monomers incorporated into each chain, and we may define
multiple subtree partial partition functions $g_M$, $M=0,1,\ldots$, in
such a way that $g_0$ corresponds to a subtree with no polymer bond
at the root, as before, and $g_M$ stands for a subtree with a polymer
bond on the 
root connected to $M$ monomers above. The recursion relations for
these partial partition functions will be:
\begin{eqnarray}
g^{\prime}_0&=&g_0^{\sigma}+ \sigma z_e
g_0^{\sigma-1}\sum_{M=1}^\infty g_M+ \nonumber \\
&& \frac{\sigma(\sigma-1)}{2}z_i g_0^{\sigma-2}\sum_{M,N=1}^\infty g_M
g_N, \\
g_1^{\prime}&=&z_e g_0^\sigma,\\
g_M^{\prime}&=&\sigma z_i g_0^{\sigma-1} g_{M-1},\;M=2,3,\ldots.
\end{eqnarray}
We proceed defining the ratios of partial partition functions
$R_M=g_M/g_0$, for $M=1,2,\ldots$. The recursion relations for the
ratios will be:
\begin{eqnarray}
R_1^{\prime}&=&\frac{z_e}{D},\\
R_M^{\prime}&=&\frac{\sigma z_i R_{M-1}}{D}, M=2,3,\ldots,
\end{eqnarray}
where
\begin{equation}
D=1+\sigma z_e\sum_{M=1}^\infty R_M +\frac{\sigma(\sigma-1)}{2} z_i
\sum_{M,N=1}^\infty R_M R_N.
\end{equation}
Inspection of these set of recursion relations leads to the following
Ansatz for the fixed point values of the ratios $R_M=\alpha^{M-1}
R_1$, where $\alpha=R_1 \sigma z_i/z_e$. The ratios $R_M$ are related
to the ratio $R$ defined above through $R=\sum_{M=1}^\infty
R_M=R_1\sum_{M=1}^\infty \alpha^{M-1}=R_1/(1-\alpha)$. Using the fixed
point value for $R$ found in Eq. (\ref{fpr}), we may then find the
values of $R_1$ 
and $\alpha$ in the thermodynamic limit. The results are:
\begin{equation}
R_1=\frac{\rho_e}{\sqrt{(\rho_e+2\rho_i)(q-\rho_e-2\rho_i)}},
\end{equation}
and
\begin{equation}
\alpha=\frac{2\rho_i}{\rho_e+2\rho_i}.
\end{equation}
Now we consider the operation of attaching $q$ subtrees to the central
site of the tree. The probability that an endpoint of a chain with $M$
monomers is located on the central site will be equal to $q z_e
R_{M-1}$, and the probability to have an endpoint of a chain located
on the central site is equal to $q z_e R$. Therefore, the probability
to find a chain with exactly $M$ monomers among all chains on the
lattice with mean molecular weight $\bar{M}$ will be given by:
\begin{eqnarray}
r_M&=&\frac{R_{M-1}}{R}=(1-\alpha)\alpha^{M-2}= \nonumber \\
&&\frac{1}{\bar{M}-1} \left( \frac{\bar{M}-2}{\bar{M}-1} \right)^{M-2}.
\end{eqnarray}
This exponential distribution of molecular weights is again
independent of the coordination number $q$ and therefore is identical
to the one obtained in \cite{snd06} for the one-dimensional case $q=2$.

\section{Final discussions and comments}
\label{fdc}
Although the equilibrium polymerization model was extensively studied
in the literature, the main focus of these studies was its critical
behavior, which occurs in the limit $K_1 \to 0$ ($z_e \to 0$), where
the mean molecular weight of the chains diverges. Here, similarly to
what was done by Dudowicz et al \cite{dfd99} for a similar model,
which essential differs from the present one only by the fact that it
is canonical with respect to the iniciator molecules, we
turn our attention to the region of parameters where the model is non
critical. The calculations performed by Dudowicz et al were in the
Flory-Huggins approximation, which for simple polymer models is
equivalent to the Bethe lattice solution. It is therefore not
surprising that boths calculations lead to an exponential distribution
of molecular weights of the chains. The model
studied by Dudowicz et al is canononical with respect to the initiator
molecules, while the present model is grand-canonical with respect to
the endpoint monomers. In the polymer limit, the
distribution of molecular weights should in general be described by the
scaling behavior \cite{s92}
\begin{equation}
r_M \approx M^{\gamma-1} \exp(-M),
\end{equation}
in the limit of small overlap, reducing to an exponential decay at
large overlaps. The critical exponent $\gamma$ is in general larger
than 1, but its classical value is 1. Since the Flory-Huggins
approximation and Bethe lattice calculations should lead to classical
exponents, the exponential decay of the distribution of molecular
weights is expected in the polymer limit $\bar{M} \to \infty$, but
nonexponential decay would be allowed for finite $\bar{M}$. We are
presently innvestigating if this actually occurs when the model is
studied using better (mean field) approximations.

Although the details of the calculations above were presented for the
restricted equilibrium polymerization model, without the presence of
one-site chains, it is easy to obtain the correpondin results for the
original model, using Eqs. (\ref{yxi}), (\ref{ze}) and (\ref{zi}). For
example, the density of sites occupied by the one-site polymers is
\begin{equation}
\rho_1=\frac{x_1}{Y}\frac{\partial Y}{\partial x_1},
\end{equation}
making $x_1=1/2$. The result is
\begin{equation}
\rho_1=\frac{K_1}{1+K_1}(1-\rho_e-\rho_i).
\end{equation}
A similar calculation may be done for the density of polymers
(including the one-site chains) $\rho_p=N_p/N$, leading to:
\begin{equation}
\rho_p=\frac{K_1}{1+K_1}\left(1+\frac{1-K_1}{2K_1}\rho_e-\rho_i,
\right) 
\end{equation}
while the density of bonds is given by:
\begin{equation}
\rho_b=\frac{\rho_e}{2}+\rho_i.
\end{equation}

\section*{Acknowledgements}
MAN acknowledges a doctoral grant from the brazilian agency CNPq and
JFS thanks the same agency for partial financial assistance.

\end{document}